%% file: main.tex
\documentclass[11pt]{article}

\usepackage[final]{acl}

\usepackage{times}
\usepackage{latexsym}

\usepackage[T1]{fontenc}

\usepackage[utf8]{inputenc}

\usepackage{microtype}

\usepackage{inconsolata}

\usepackage{graphicx}

\usepackage{tabularx} 
\usepackage{booktabs}

\usepackage[table]{xcolor}
\usepackage{colortbl}
\usepackage{multirow}

\usepackage{float}

\usepackage{enumitem}

\title{In-Browser Human-AI Co-Writing for Counterspeech: An Exploratory User Study of Agency, Perceived Response Quality, and Time}

\author{
  Tammy Schmidt \and
  Luise Koch\\
  Technical University of Munich\\
    \small{\href{mailto:tammy.schmidt@tum.de}{tammy.schmidt@tum.de}, \href{mailto:luise.koch@tum.de}{luise.koch@tum.de}}
}

\begin{document}
\maketitle
\input{sections/00_Abstract}

\input{sections/01_Introduction}
\input{sections/02_RelatedWork}
\input{sections/03_System}

\input{sections/04_Experiments}

\input{sections/05_Discussion}
\input{sections/06_Conclusion}

\input{sections/07_Limitations}

\section*{AI Assistance Disclosure}

The authors used generative AI tools, including OpenAI (GPT-4/o), Google Gemini and Claude to support language editing, shortening, structural revision, and code refinement. All generated suggestions were critically reviewed, verified, and revised by the authors, who take full responsibility for the submitted content.

\section*{Acknowledgments}
We thank all participants who took part in our study for their time and valuable contributions.


\bibliography{custom}

\appendix
\setcounter{figure}{0}
\setcounter{table}{0}

\renewcommand{\thefigure}{A.\arabic{figure}}
\renewcommand{\thetable}{A.\arabic{table}}

\input{sections/A_Appendix}

\end{document}

%% file: sections/00_Abstract.tex
\begin{abstract}
Writing counterspeech that challenges online hate speech can be cognitively and emotionally demanding, whereas fully automated generation may weaken users’ sense of voice and responsibility. We present an in-browser AI-assisted co-writing tool that enables users to contribute their own ideas and offers three editable suggestions generated by a large language model using dynamically retrieved hate speech–counterspeech examples from the Multitarget CONAN dataset. We evaluated the tool in an exploratory within-subject study with ten participants, complemented by think-aloud observations and semi-structured interviews with half of the participants. Participants wrote responses to three self-selected hate speech instances without assistance and then repeated the task with the tool; final responses were subsequently peer-rated for empathy, non-toxicity, relevance, specificity, and persuasiveness. Participants evaluated the interface positively, and qualitative accounts linked user input, selection among alternatives, and small edits to perceived agency and responsibility. Assisted responses received higher descriptive mean ratings across all five proxies, with the largest difference for empathy. Mean composition time increased from 2.52 to 3.88 minutes. Qualitative findings further revealed a tension between empathy-oriented outputs and users’ desire to express legitimate anger or direct disagreement. We contribute exploratory design guidance for treating AI-assisted counterspeech as deliberative co-writing rather than automated text production.
\end{abstract}

%% file: sections/01_Introduction.tex
\section{Introduction}\label{chapter:introduction}

Online hate speech (HS) is a persistent feature of digital communication that can harm both its direct targets and the wider communities exposed to it. Prior research links exposure to HS to psychological distress, reduced feelings of safety, and the normalization of hostile discourse \citep{cohen-almagorFightingHateBigotry2011, UnsocialMediaReal2017, hindujaBullyingCyberbullyingSuicide2010}. Platforms commonly respond through content removal or account sanctions, but such measures remain contested and do not enable users to participate constructively in addressing harmful content. Counterspeech (CS) offers an alternative by challenging hateful narratives through non-aggressive responses that support targets, encourage reflection, and model more civil forms of online interaction for bystanders \cite{beneschCounteringDangerousSpeech2014, schiebGoverningHateSpeech2016, baiderAccountabilityIssuesOnline2023}. However, many users refrain from engaging because composing a response can be time-consuming, emotionally demanding, and rhetorically difficult \cite{munCounterspeakersPerspectivesUnveiling2024, pingCounterExploringMotivations2025}. 

Large language models (LLMs) can generate fluent and contextually relevant CS, particularly when adapted through prompting or fine-tuning \cite{dogancGenericPersonalizedInvestigating2023, zhengWhatMakesGood2023}. However, most work evaluates generated outputs or model configurations, leaving comparatively less evidence about how artificial intelligence (AI) should be embedded in interactive tools that support users without replacing their judgment, voice, or responsibility \cite{munCounterspeakersPerspectivesUnveiling2024, pingCounterExploringMotivations2025}. This interaction question is important since a CS comment represents a public moral and social intervention.

We address this gap through the design and exploratory evaluation of an in-browser CS co-writing assistant. The interface requires user input, presents three editable LLM-generated CS suggestions, and leaves selection, revision, and posting entirely with the user. We contribute: (1) an interaction design that operationalizes user initiation, contribution, choice, and editing; (2) exploratory evidence on users’ experiences with the tool and descriptive differences between assisted and unassisted responses and subsequent peer evaluation; and (3) design implications concerning agency, rhetorical diversity, composition time, and the tension between empathy-oriented safeguards and personal voice. 

\noindent
We examine the following research questions:\\

\textbf{RQ1}: How do users experience an in-browser AI-assisted CS co-writing interface in terms of usability, agency, perceived ownership, and personal voice? \\

\textbf{RQ2}: How do assisted and unassisted responses differ descriptively in peer-rated response-quality proxies and composition time? \\

%% file: sections/02_RelatedWork.tex
\section{Related Work}\label{chapter:relatedwork}

\subsection{Counterspeech Strategies and Evaluation}
CS effectiveness can refer to different outcomes, including changing a hate speaker’s stance, de-escalating an exchange, supporting targets, encouraging bystander intervention, or improving the tone of a wider discussion \cite{baiderAccountabilityIssuesOnline2023, buergerCounterspeechLiteratureReview2021}. Research on pro-social example setting suggests that visible civil responses can influence subsequent participation \cite{hanCivilityContagiousExamining2018, seeringShapingProAntiSocial2017}. However, strategy effects depend on context and audience. Constructive, empathetic, and non-hostile responses are often evaluated favorably \cite{baiderAccountabilityIssuesOnline2023, fraserWhatMakesGood2023, hangartnerEmpathybasedCounterspeechCan2021}, whereas findings for humor, questioning, and denunciation are mixed \cite{fraserWhatMakesGood2023, beneschConsiderationsSuccessfulCounterspeech}. AI assistance may help reduce the emotional burden of formulating CS after encountering HS and support users in crafting measured, well-formulated responses \cite{fraserWhatMakesGood2023, pingCounterExploringMotivations2025}. Accordingly, we treat empathy, non-toxicity, relevance, specificity, and persuasiveness as perceived linguistic qualities rather than as direct measures of real-world CS effectiveness.

\subsection{Automatic Counterspeech Generation} \label{sec:RW_CSgeneration}

Automated CS generation has been approached through fine-tuning \cite{tekirogluUsingPreTrainedLanguage2022, zhengWhatMakesGood2023, zhuGeneratePruneSelect2021}, prompt engineering \cite{ashidaAutomaticGenerationMessages2022, zhengWhatMakesGood2023, lyuHWTSCMultilingualCounterspeech2025}, and retrieval-augmented generation (RAG) \cite{leekhaWarWordsHarnessing2024, chungKnowledgeGroundedCounterNarrative2021}. No method is universally superior, although prompting large pretrained models has outperformed smaller fine-tuned models in some CS settings \cite{dogancGenericPersonalizedInvestigating2023, zhengWhatMakesGood2023}. In this context, and with the complementary goal of lowering computational requirements, we use retrieval-augmented few-shot prompting. Prior work used static Multitarget CONAN examples for in-context learning in CS generation tasks \cite{ashidaAutomaticGenerationMessages2022}. The present system extends this approach by dynamically retrieving different examples for each prompt based on semantic similarity. This design may increase contextual alignment, but the present study focuses on human–AI interaction and does not independently test whether dynamic retrieval improves generation quality.\\

\subsection{Interactive Tools for Counterspeech}\label{sec:RW_tools}

Much CS research focuses on optimizing generation models, while paying less attention to how users interact with AI assistance during the writing process. Prior studies therefore argue that AI-based CS systems should support human judgment rather than operate as autonomous agents \cite{munCounterspeakersPerspectivesUnveiling2024, pingCounterExploringMotivations2025, huangEmpoweringCreatorsFight2025}. This assistive framing is important because users seek help with formulation, emotional regulation, and factual support, but also raise concerns about agency, authenticity, and responsibility \cite{munCounterspeakersPerspectivesUnveiling2024}.\\
\indent Two systems are particularly relevant. 
\citet{chungEmpoweringNGOsCountering2021} support practitioners responding to Islamophobic HS by monitoring trends and retrieving pre-written responses from the CONAN dataset, but their system does not generate new, input-specific CS. \textit{CounterQuill} combines education, brainstorming, and AI-assisted co-writing to support skill development and perceived authorship \citep{dingDesigningHumanAICollaboration2025}. Its multi-step workflow offers substantial guidance but requires users to leave the immediate context of the abusive content. Our prototype explores a lower-friction alternative by placing co-writing support inside the browser and emphasizing user input, choice, and editing. Recent browser-based work has explored expert-grounded RAG support for German CS in moderation contexts \citep{filliesCounteringGermanHate2026}. Our prototype complements this line of work by studying an in-browser co-writing assistant for individual users, emphasizing substantive user input, choice among three editable LLM-generated suggestions, and perceived agency in the writing process.

\raggedbottom

%% file: sections/03_System.tex
\section{System}\label{chapter:design}

\subsection{Interaction Design} \label{sec:dev_designspace}

Drawing from the design space for intelligent writing assistants proposed by \citet{leeDesignSpaceIntelligent2024}, the extension is designed as a co-writing tool rather than an autonomous CS generator. All interaction choices are intended to preserve users' agency, perceived ownership, and responsibility for the final message, particularly through explicit system steering, selection among alternatives, and post-editing.

\subsection{Ethics-by-Design} \label{sec:dev_ethics}

Because the system processes hateful content and potentially sensitive user input, ethical considerations informed its design from the outset. The design process was guided by the \textit{Eight principles of ethical AI} synthesized by the Berkman Klein Center for Internet \& Society at Harvard University \cite{fjeldPrincipledArtificialIntelligence2020}. In particular, the extension operates only upon explicit user initiation, minimizes the data it collects, and leaves decision-making power entirely with the user. Users are informed that their input is transferred to an externally hosted LLM and must consent before accessing the generator. Prompt-level safeguards aim to reduce harmful or toxic output, although reliance on a pretrained model and an external API limits the developers' control over model behavior and downstream data processing. A detailed account of the technical and organizational measures is provided in Table \ref{tab:ethics}.\\

\subsection{System Architecture} \label{sec:dev_architecture}

\begin{figure*}[t]
    \centering
    \includegraphics[width=\textwidth]{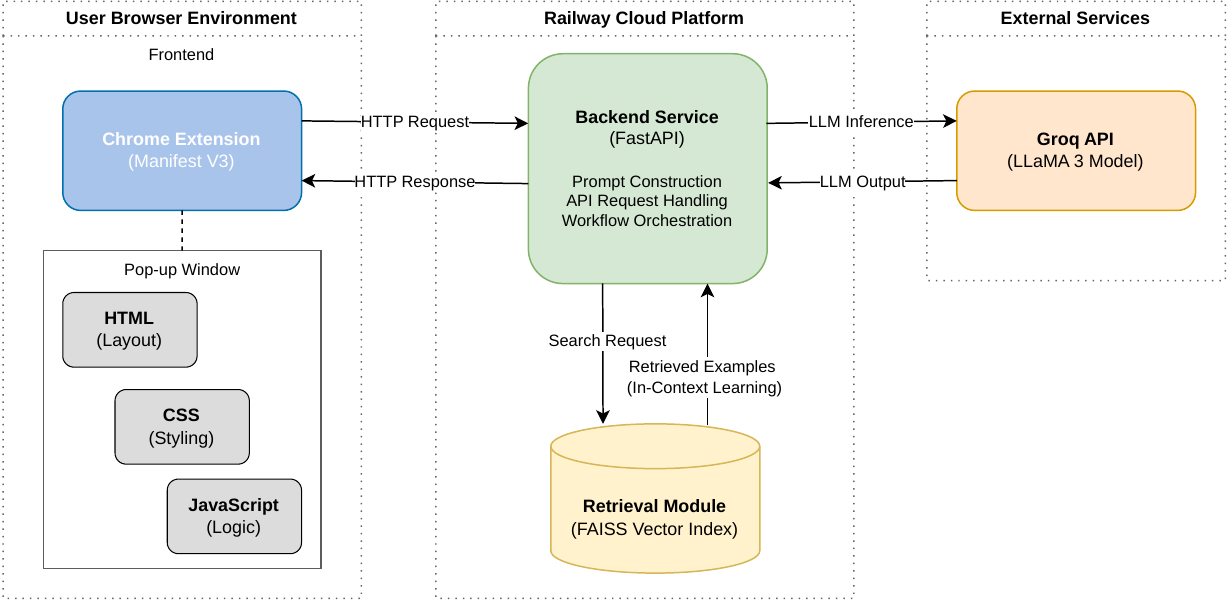}
    \caption{System architecture diagram.}
    \label{fig:architecture}
\end{figure*}

The system follows a modular client–server architecture consisting of a browser-based frontend, a backend service, a retrieval module, and an externally hosted LLM. Figure \ref{fig:architecture} provides an overview of the system architecture.\\

\paragraph{Frontend}

The frontend is implemented as a Manifest V3 browser extension with a compact pop-up interface. It contains two views: (1) the Generator window and (2) the Info page. 

In the Generator window, users enter the hateful comment and must provide their own ideas, wording, or personal perspective in a free-text field. This mandatory contribution is intended to ground the generated responses in users' own views rather than encourage passive acceptance of fully automated text. Users additionally specify their responder role in relation to the HS target group (direct target, target-group member, or ally), select a writing style (familiar or formal), and choose an approximate response length (short, medium, or long). The system then presents three editable CS suggestions. Users can compare the alternatives, select one, revise the text, and copy the final response.
Screenshots are provided in Figures \ref{fig:pop-up_generator_page}, \ref{fig:filled-out_generator_page} and \ref{fig:suggestions_generator_page}. 

The Info page explains data processing, user accountability, and appropriate use. Users must acknowledge these conditions before accessing the Generator. The complete consent statements are shown in Figure \ref{fig:consent_statements}. \\

\paragraph{Backend and Prompt Construction} 

The backend is implemented in Python using FastAPI. For each request, it validates the submitted inputs, retrieves examples for in-context learning, constructs the model messages, and sends them to the LLM through the Groq API. The system message specifies the CS task, response-quality guidelines, and safeguards. The prompt message contains the HS instance, the user's contribution and selected controls, the retrieved examples, and the required output format. Prompt templates are provided in Tables \ref{tab:system_message} and \ref{tab:prompt_content}. The returned output is parsed into three separate suggestions and sent to the frontend. \\

\paragraph{Example Retrieval} 

The retrieval component uses the Multitarget CONAN dataset, which contains 5,003 HS-CS pairs covering eight target categories: disabled people, Jewish people, LGBT+ people, migrants, Muslims, People of Color, women, and others \cite{fantonHumanintheLoopDataCollection2021}. It was selected because of its size, target-group coverage, and the expert review and post-editing involved in its construction.

The HS instances are embedded using \textit{all-mpnet-base-v2} and stored in a FAISS vector index. At generation time, the submitted HS is embedded and matched against the index using k-nearest-neighbor search. The five most semantically similar HS-CS pairs are added to the prompt as dynamically selected in-context demonstrations. We fixed the number of demonstrations at five as an implementation choice; the effect of varying this parameter was not evaluated. \\

\paragraph{Large Language Model Integration}

The writing assistant uses \texttt{llama-3.3-70b-versatile}, a model from the Llama~3 family \cite{grattafioriLlama3Herd2024}, accessed through the Groq API\footnote{\url{https://console.groq.com/home}}. The availability of model documentation and open weights provides partial transparency, although the model remains externally hosted and was not modified in this study. The model was invoked with a temperature of 1.0, a generation setting that encourages more varied outputs, following prior CS generation work that used higher temperature settings to encourage output diversity \cite{lyuHWTSCMultilingualCounterspeech2025, dogancGenericPersonalizedInvestigating2023}; provider defaults were used for all other parameters.

Each prompt component was designed to play a distinct role. The user contribution was intended to ground the response in the user’s ideas or wording, while the interface controls specified the responder role, style, and approximate length. The retrieved examples served as semantically related in-context demonstrations. In contrast, the system message imposed default response-quality guidelines and safeguards. Importantly, the prompt’s priority order placed safeguards and default guidelines above user input, with retrieved examples receiving the lowest priority. These roles describe the intended prompt design rather than measured component effects. Because the components were not manipulated independently, the present study cannot determine whether topical relevance arose from retrieval, prompt construction, or model tendencies.

%% file: sections/04_Experiments.tex
\section{Experiments} \label{chapter:experiments}

\input{sections/04_01_ExperimentalSetup}

\input{sections/04_02_EvaluationMethods}

\input{sections/04_03_Results}

%% file: sections/04_01_ExperimentalSetup.tex
\subsection{Experimental Setup} \label{sec:exp_setup}

\paragraph{Study Design}

The user study used an exploratory, fixed-order within-subject design followed by a peer-rating comparison. Participants were presented with the same set of eight HS instances from the Multitarget CONAN dataset, comprising one instance for each target category (see Table \ref{tab:HS_instances_userstudy}). Instances were manually screened to ensure suitability and variation in hate intensity. Exact duplicates of the study stimuli were removed from the FAISS index to prevent exact-match retrieval during generation; closely related examples nevertheless remained.

In the first round, participants freely selected three out of the eight HS instances and manually wrote a CS response to each of them. In the second round, they responded to the same instances using the browser extension for assistance. Holding the selected stimuli constant within participants enabled a descriptive comparison of their assisted and unassisted responses, although the fixed order and repeated exposure may have introduced practice, memory, or demand effects. Because participants selected their own stimuli, instances were not represented equally across the sample. 
After each unassisted response, participants rated their willingness to post the text publicly under their own identity. In the assisted condition, participants rated each final response on five exploratory items: self-identity, sense of agency, accountability, naturalness, and willingness to post. Willingness to post was therefore the only item assessed in both conditions; the other four items specifically concerned participants’ relationship to AI-assisted co-written text.

The study was conducted in two batches of five participants. The first batch additionally completed think-aloud observations during task completion and took part in a short semi-structured interview afterwards; the second batch completed the same writing and rating procedure without the additional qualitative component. No saturation criterion was applied, as this qualitative component was intended to provide illustrative rather than comprehensive thematic coverage. The first author reviewed recordings, extracted relevant passages, and grouped them inductively into broad recurring themes. No predefined coding scheme or independent second coding was used. Qualitative observations are therefore used to contextualize descriptive ratings and identify possible design implications. Quotations were lightly edited to remove filler words or correct grammatical inconsistencies for improved readability while strictly preserving meaning. 

About a week later, participants within each batch peer-rated responses written by others on five perceived response-quality dimensions. Self-ratings were excluded to reduce self-evaluation bias, resulting in four peer-ratings for each response. Evaluation concerned the final responses, including any edits made by participants, assessing the resulting human or human–AI-produced text rather than the isolated performance of the underlying model.\\

\paragraph{Participants}\label{sub:participants}

The study included ten adults aged 23 to 32 (M$= 27.5$) with English proficiency; five identified as women and five as men. The sample included participants from six national backgrounds and participants with LGBTQ+, Muslim, migration, and racialized identities. Three participants had previously written CS on a small number of occasions, while the remaining participants reported no prior experience. Two participants had professional research experience related to online hate mitigation. Participants were recruited through the authors' personal and professional networks. Participation was voluntary and uncompensated. \\

\paragraph{Ethical Considerations}\label{sub:ethicalconsiderations}

Because the study involved exposure to potentially offensive and distressing content, several measures were taken to reduce risks to participants. All participants received information about the study procedures and provided informed consent before taking part, including consent for the think-aloud and interview recordings. During recruitment and again before each study phase, they were warned about the nature of the material and the groups targeted by the HS instances. Participants were informed that they could withdraw from the study at any time without providing a reason and without negative consequences, and contact information was provided if they experienced distress related to the study. Study data were stored securely, and access was restricted to the research team. The study did not undergo formal institutional ethics review.

%% file: sections/04_02_EvaluationMethods.tex
\subsection{Evaluation Methods}

\paragraph{User Experience}

User experience with the browser extension was assessed using the Short Version of the User Experience Questionnaire (UEQ-S)\footnote{\url{https://www.ueq-online.org/}}. The UEQ-S comprises eight bipolar adjective pairs rated on a seven-point scale from $-3$ to $3$. Four items measure pragmatic quality, addressing task-related support, while four measure hedonic quality, measuring stimulation and novelty \cite{schreppDesignEvaluationShort2017}. The short version was selected to limit participant burden at the end of the study while retaining a standardized measure of interaction experience.

The UEQ-S benchmark provides reference values for positioning ratings relative to other interactive products \cite{hinderksBenchmarkShortVersion2018}. Given the sample size of ten participants, however, benchmark comparison is interpreted only as an exploratory indication rather than as evidence of robust comparative performance. UEQ-S results are used primarily to identify initial strengths and areas for refinement in the early-stage prototype. \\

\paragraph{Authenticity-Related Perceptions}

To examine users' relationship to co-written responses, the study included five exploratory items addressing self-identity, sense of agency, accountability, naturalness, and willingness to post. These dimensions were informed by prior work discussing authenticity, agency, and authorship in AI-assisted CS writing, for which no standardized measurement approach has yet been established \cite{munCounterspeakersPerspectivesUnveiling2024, pingCounterExploringMotivations2025, huangEmpoweringCreatorsFight2025, dingDesigningHumanAICollaboration2025}. Participants rated assisted responses on all five items using a five-point Likert scale. The exact wording is provided in Table~\ref{tab:authenticity_likert}. The items are interpreted separately rather than as a unified or validated authenticity scale. \\

\paragraph{Peer Rated Response-Quality Proxies}

Human evaluation was chosen because automatic generation metrics frequently correlate only weakly or inconsistently with human judgments of response quality and preference \cite{mousaviEvaluationResponseGeneration2022, qianBenchmarkDatasetLearning2019}. Peer raters evaluated the (co-)written CS with regard to five textual qualities commonly associated with constructive CS: empathy \cite{qianHarnessingPowerLarge2023, leeDoesGPT3Generate2022}, non-toxicity \cite{ashidaAutomaticGenerationMessages2022, hengleIntentconditionedNontoxicCounterspeech2024}, relevance \cite{jonesMultiAspectFrameworkCounter2024a, zhuGeneratePruneSelect2021}, specificity \cite{jonesMultiAspectFrameworkCounter2024a}, and persuasiveness \cite{wangIntentAwareHateMitigatingCounterspeech2024}. Definitions as presented to raters are provided in Table~\ref{tab:effectiveness_likert}. Each HS–CS pair was shown in randomized order and rated on a five-point Likert scale. The ratings are interpreted as perceived response-quality proxies rather than as measures of real-world effectiveness. 

Ordinal Krippendorff's $\alpha$ \cite{krippendorffContentAnalysisIntroduction2018} was calculated separately for each criterion to assess agreement among the four peer raters. Values of $1$ indicate perfect agreement, while values around $0$ indicate agreement no greater than expected by chance; negative values indicate systematic disagreement. \\

\paragraph{Composition Time}

Composition time was recorded in seconds for each response in both study rounds. Based on prior work suggesting that AI assistance may reduce the effort required to formulate CS \cite{chungEmpoweringNGOsCountering2021}, we expected shorter composition times in the assisted condition than in the unassisted condition. Longer duration may reflect additional interface interaction, comparison of alternatives, or editing; these activities were not measured separately. Composition time is therefore reported descriptively, without assuming that shorter or longer completion times are inherently preferable.

%% file: sections/04_03_Results.tex
\subsection{Results}

\paragraph{User Experience}

In the UEQ-S evaluation, pragmatic quality received a mean rating of $2.2$ ($SD = 0.6$), hedonic quality a mean of $1.7$ ($SD = 1.2$), and the overall scale a mean of $1.9$ ($SD = 0.8$). These descriptive results indicate that participants generally evaluated the prototype positively. A detailed overview of the ratings can be found in Table~\ref{tab:UEQ-S_per_item}; aggregated values are mapped against the UEQ-S
benchmark in Figure~\ref{fig:UEQ-S_Excellent}. However, given the sample size, results should be interpreted as an initial assessment of the interaction experience rather than as a robust comparison with other products. 

Qualitative feedback from the five think-aloud interview participants was broadly consistent with these ratings. Participants described the interface as intuitive and appreciated its availability within the browser and its tailoring to CS writing. One participant explained:

\begin{quote}
\textit{``I think it's a very useful tool, especially within the browser. Maybe other people would go over to ChatGPT or some other LLM and would look for something there. So you already have a fairly controlled area that is tailored to it. I think that's pretty good.''}
\end{quote}

Participants also identified opportunities for refinement. In particular, they requested more iterative editing options, greater transparency, and support for factual arguments. Some participants further questioned whether they would engage in CS outside the study, suggesting that assistance with response formulation alone may not address users' broader motivation to intervene. \\

\paragraph{Authenticity-Related Perceptions}

Participant-level mean willingness to post ratings -- the only criterion rated under both conditions -- increased for (n = 4), decreased for (n = 2), and remained unchanged for (n = 4), indicating no clear directional trend or consistent effect at the individual level.

\begin{figure*}[t]
    \centering
    \includegraphics[width=\textwidth]{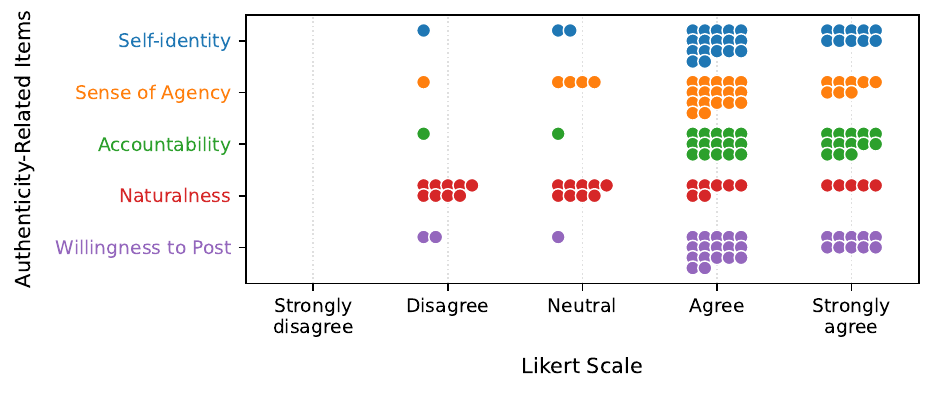}
    \caption{Distribution of Likert scale ratings per authenticity-related item for CS written with assistance of the browser extension.}
    \label{fig:dist_authentic}
\end{figure*}

Figure~\ref{fig:dist_authentic} shows the distribution of all five exploratory ratings collected for responses composed with the assistance of the browser extension. Each item comprises 30 ratings, corresponding to three assisted responses from each of the ten participants. Ratings were generally concentrated in the “agree” and “strongly agree” categories, particularly for self-identity, accountability, and willingness to post. Naturalness received the least favorable distribution.

Qualitative feedback helps contextualize the lower naturalness ratings. Participants frequently described the suggestions as generic and observed that requests for sarcastic, witty, judgmental, or fact-based responses were often not reflected in the output. Participants perceived the system as favoring empathetic formulations, which sometimes conflicted with how they wished to respond. One participant stated:

\begin{quote}
\textit{``I do not like these textbook phrases, `I understand where you're coming from.' No, I don't understand. So those suggestions do not appeal as much to me.''}
\end{quote}

All five qualitative participants requested a broader range of response styles. Their feedback suggests a tension between producing consistently de-escalating responses and supporting users in expressing their own emotional position.

Participants also indicated that contributing their own ideas, selecting among three suggestions, and revising a selected response were important to their sense of agency and responsibility. One participant reflected:

\begin{quote}
\textit{``I did feel like I'm the one carrying the whole responsibility for that statement because it was based on my idea. But I knew that I'm not the one actively writing every single sentence.''}
\end{quote}

These observations are illustrative, since no systematic interaction logs were collected to compare selection and editing behavior. \\

\paragraph{Peer Rated Response-Quality Proxies} 

Table \ref{tab:results_effect} reports the descriptive mean peer ratings for the unassisted and assisted responses. Mean ratings were higher in the assisted condition across all five quality criteria, with differences ranging from $0.19$ for relevance to $0.88$ for empathy.

Qualitative feedback was consistent with the higher empathy ratings. Several participants described their unassisted responses as sharper or more confrontational and perceived the generated suggestions as more restrained. One participant explained: 

\begin{quote}
\textit{``When I go to social media to reply to someone, I'm not really nice. And maybe that way I position myself as a member of the [target] group in the wrong light. So it would help people like me to have a nicer way of responding to this.''}
\end{quote}

At the same time, participants frequently characterized the generated responses as generic. This contrasts with the slightly higher mean peer rating for specificity in the assisted condition and illustrates that external assessments of textual specificity did not necessarily correspond to authors' expectations of a distinctive response.

Inter-rater agreement was low across all criteria, with Krippendorff's $\alpha$ ranging from $-0.019$ to $0.314$ across batches and response-quality proxies; complete results are provided in Table~\ref{tab:iaa_results}. Condition means should therefore be interpreted cautiously as descriptive patterns rather than reliable evidence of improved response quality. \\

\begin{table}[]
    \centering
    \begin{tabular}{lccccc}
        \hline
        \textbf{Round} & EMP & NTOX & REL & SPEC & PERS \\
        \hline
        \textbf{1} & 3.11 & 3.87 & 4.09 & 3.69 & 3.29 \\
        \textbf{2} & 3.98 & 4.20 & 4.28 & 3.90 & 3.60 \\
        \hline
        \textbf{$\Delta$} & 0.87 & 0.33 & 0.19 & 0.21 & 0.31 \\
        \hline
    \end{tabular}
    \caption{Mean peer ratings of five response-quality proxies for (1) unassisted and (2) assisted CS responses (30 responses per condition; four ratings per response).}
    \label{tab:results_effect}
\end{table}

\paragraph{Composition time} 

Nine of the ten participants took longer when using the browser extension. Contrary to our initial expectation, mean composition time increased from 2.52 minutes in the unassisted condition to 3.88 minutes in the assisted condition, corresponding to a mean difference of 1.36 minutes, or approximately 54\%. Participant-level mean composition times are shown in Figure ~\ref{fig:time_per_round}. Thus, assistance did not reduce composition time in this study.\\

%% file: sections/05_Discussion.tex
\section{Discussion}\label{chapter:discussion}

The study provides exploratory evidence about how an in-browser AI assistant can shape CS writing, particularly with respect to agency, rhetorical style, perceived response quality, and composition time.

\paragraph{Interaction Design}

Participants valued the low-friction browser interface, personal input, and choice among multiple CS suggestions. Their accounts suggest that agency did not require composing every sentence from scratch as long as they retained substantive and editorial control. However, we did not collect data on participants' prior use of LLMs, so familiarity with AI writing tools may have shaped perceptions of agency and control. Since post-generation editing was limited, future CS tools should support more iterative revision at the phrase or argument level, for example, by allowing users to request alternatives for selected spans or save suggestions for comparison.

\paragraph{De-Escalation and Authenticity}

The strongest conceptual tension concerned empathy, which was encoded as an explicit default guideline and placed above potentially conflicting user instructions in the prompt hierarchy. In line with this design choice, assisted responses received the largest descriptive increase in peer-rated empathy. Some participants valued the system's ability to temper anger, while others -- particularly those who identified with the targeted group -- experienced formulaic empathy toward the hate speaker as inauthentic or invalidating. This cautions against treating empathy as a universal design objective or proxy for effectiveness. Although the interface allowed users to specify role, register, and approximate length, it did not let them explicitly choose among rhetorical CS strategies. Future systems could offer several non-toxic strategies, such as factual correction, norm-based disapproval, questioning, solidarity with targets, boundary setting, or restrained humor.

User satisfaction may also depend on the quality of the LLM suggestions and on the specific prompt configuration used in this study. Future ablation studies should therefore vary components such as interface controls, system guidelines, and retrieved examples to assess their respective influence on output quality and perceived usefulness.

\paragraph{Perceived Response Quality and Time}

All five mean peer ratings were higher for assisted responses. However, the fixed-order design prevents us from attributing these differences to the extension alone, since participants may have benefited from practice, repeated exposure to the same stimuli, or expectations about the study purpose. Low and sometimes negative inter-rater agreement further indicates that raters interpreted the evaluation criteria inconsistently. Future evaluations could therefore examine not only average ratings, but also whose judgments differ and why, for example, through perspectivist analyses \cite{basilePerspectivistTurnGround2023}.

The extension increased mean composition time by 54\%, contrary to the initial efficiency expectation. Longer time may reflect comparison of suggestions, experimentation, or revision, but the study did not record sufficiently fine-grained interaction logs to determine the mechanism. Importantly, longer composition time does not necessarily imply lower CS participation overall: lower access friction could still increase whether users begin, complete, or post a response. Future field or choice-based studies should therefore measure response initiation and posting behavior alongside time, perceived efficacy, and response quality. The prototype should be understood as a deliberative co-writing aid rather than a demonstrated time-saving tool.

%% file: sections/06_Conclusion.tex
\section{Conclusion}\label{chapter:conclusion}

We presented an in-browser AI-assisted tool for CS co-writing and explored its use with ten participants, focusing on how users experience such an interface and how assisted and unassisted responses compare descriptively. Regarding RQ1, participants evaluated the interface positively, while qualitative accounts associated substantive user input, choice among alternatives, and editing with perceived agency and responsibility. However, generic and strongly empathy-oriented suggestions sometimes conflicted with users’ personal voice, motivating support for a broader range of non-toxic rhetorical strategies. Regarding RQ2, assisted responses received higher descriptive mean peer ratings across all five quality proxies, but low inter-rater agreement and the fixed-order design preclude causal interpretation; composition time also increased. Together, these findings position AI-assisted CS as a deliberative co-writing process rather than a shortcut to automated text production and offer exploratory design guidance for tools that support users in crafting constructive, personally accountable responses to online hate while preserving their agency and voice.\\

%% file: sections/07_Limitations.tex
\section*{Limitations}\label{chapter:limitations}
The study’s small convenience sample of ten participants limits the robustness and transferability of the findings. Participants were recruited through the authors’ networks and were all English-proficient adults, with two having professional experience with online hate mitigation. Only half of the sample contributed think-aloud and interview data. The results may therefore be shaped by a relatively engaged and technology-comfortable participant group and may not fully capture the perspectives of broader user populations, professional counterspeakers, or users from different linguistic and cultural contexts.

The within-subject procedure used a fixed unassisted-then-assisted order and repeated the same stimuli. It therefore cannot separate the extension from practice, memory, fatigue, demand, or order effects. Participants also selected three of eight stimuli, creating non-random exposure and reducing comparability across individuals. Both the stimuli and retrieved examples came from the semi-automatically constructed, expert-reviewed CONAN dataset rather than from live platform conversations. They therefore lacked surrounding interaction history, platform cues, audience information, and the social context in which CS would ordinarily be written. Therefore, the results may not generalize to novel target groups, platform-specific language, implicit or coded HS, or multi-turn conversational contexts. Observation by the researcher may additionally have increased socially desirable responses. No participant posted CS in a live environment, so the study does not measure uptake, target support, bystander response, hate speaker behavior, or other real-world outcomes. The results therefore support provisional interaction-design insights, not causal or generalizable claims that the extension improves CS effectiveness.

Finally, the study evaluated the complete prototype rather than individual technical components. It cannot attribute outputs to the model, prompt, safeguards, retrieved examples, temperature, interface controls, or external service, as these variables were neither independently controlled nor optimized in the present study. Consequently, the reported findings should be interpreted as evidence for the viability of the overall system rather than as an assessment of any single design choice, and results may differ substantially depending on how these components are configured or combined. \\

%% file: sections/A_Appendix.tex
\section{Appendix}

\input{sections/A1_System}
\input{sections/A2_Evaluation}

\input{sections/A3_Results}

%% file: sections/A1_System.tex
\textbf{Offensive Content Warning} 

The appendix includes screenshots and user study examples that contain hateful comments directed toward the following groups: people with disabilities, Jewish people, LGBTQ+ people, migrants, Muslims, People of Color, women, and others. This is strictly for the purposes of enabling this research. Please be aware that this content could be offensive and cause you distress. \\

\begin{figure*}[ht]
    \centering
    \begin{minipage}{0.48\textwidth}
        \centering
        \includegraphics[width=\linewidth]{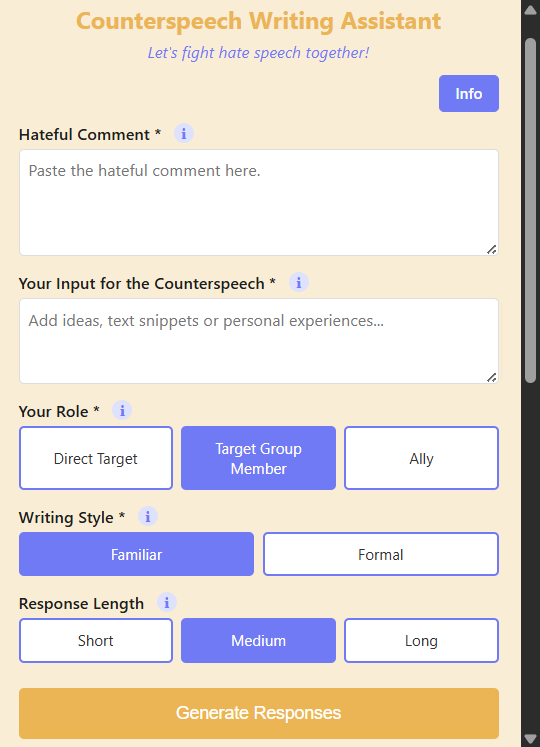}
        \caption{Screenshot of the default Generator page of the browser extension's pop-up window.}
        \label{fig:pop-up_generator_page}
    \end{minipage}
    \hfill
    \begin{minipage}{0.48\textwidth}
        \centering
        \includegraphics[width=\linewidth]{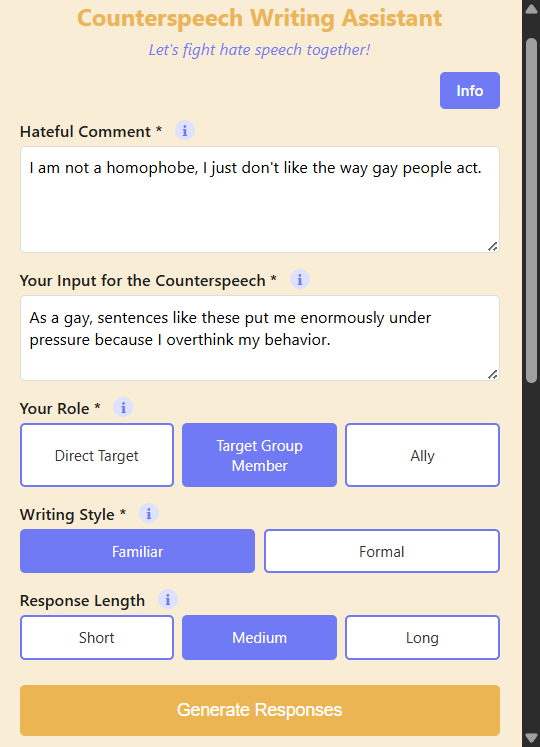}
        \caption{OFFENSIVE CONTENT WARNING: Screenshot of sample filled out Generator page.}
        \label{fig:filled-out_generator_page}
    \end{minipage}
\end{figure*}

\begin{figure*}[ht]
    \centering
    \begin{minipage}{0.48\textwidth}
        \centering
        \includegraphics[width=\linewidth]{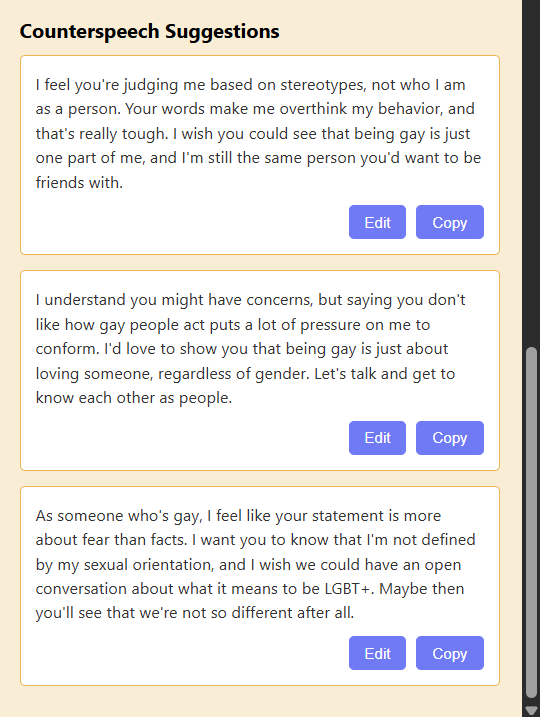}
        \caption{Screenshot of sample generated CS suggestions.}
        \label{fig:suggestions_generator_page}
    \end{minipage}
    \hfill
    \begin{minipage}{0.48\textwidth}
        \centering
        \includegraphics[width=\linewidth]{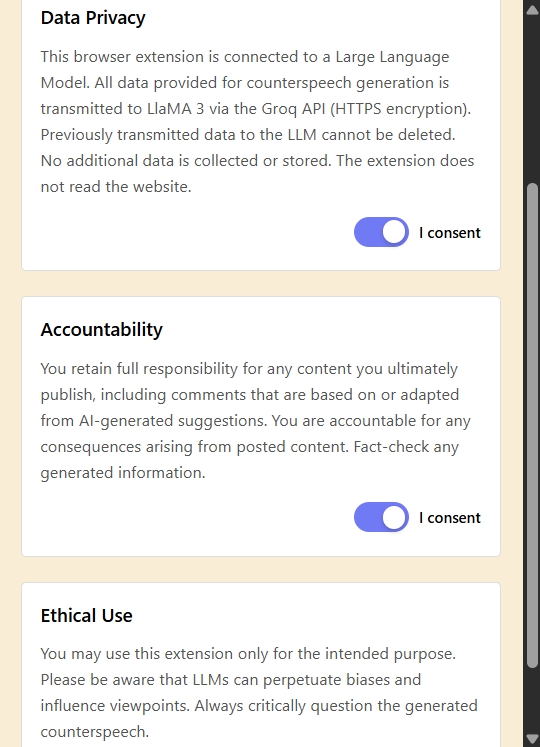}
        \caption{Consent statements on the browser extension's Info page.}
        \label{fig:consent_statements}
    \end{minipage}
\end{figure*}

\begin{table*}[ht]
\centering
    \begin{tabular}{>{\bfseries}p{2.6cm} p{5.9cm} p{5.9cm}}
        \toprule
        Principle & Implementation & Limitations \& Trade-offs \\
        \hline
        
        Privacy &
        No DOM scanning; only processes comments on explicit user request. No identifying data collected; user consent required for data transfer to the LLM. No tracking or persistent identifiers. Server hosted within the EU (Amsterdam). Dataset sourced from expert-authored content, not online users. &
        Data transmitted to the Groq API may be processed outside the EU. Previously transmitted data cannot be retroactively deleted. \\
        \hline
        
        Accountability &
        Users retain full responsibility for published content, explicitly acknowledged during installation. Human oversight is maintained throughout. &
        The underlying LLM is an external service; training data, internal processes, and output behavior are beyond direct control. \\
        \hline
        
        Safety \& Security &
        API credentials stored as environment variables; never exposed publicly. All external communication encrypted via HTTPS. Prompt-level guardrails instruct the model to detect and reject harmful user input \citep{dingDesigningHumanAICollaboration2025}. &
        The persuasive nature of CS introduces a risk of misuse to reinforce rather than counter harmful viewpoints. \\
        \hline
        
        Transparency \& Explainability &
        Users are informed about system operation, model identity, and data transmission after installation. The availability of Llama’s model documentation and open weights supports transparency. &
        Explainability of individual outputs is of limited applicability given the tool's assistive (non-autonomous) role. \\
        \hline
        
        Fairness \& Non-discrimination &
        No demographic data collected; safeguards aim to prevent generation of toxic or discriminatory content. Users are informed that outputs may reflect model biases. &
        LLMs including LLaMA can reproduce social biases from training data \citep{touvronLLaMAOpenEfficient2023, ranaldiTripFairnessBias2024}. Processing abusive input may reinforce the model's capacity to reproduce harmful language. \\
        \hline
        
        Human Control &
        All CS suggestions are presented for user review, modification, and post-editing before publication. Users retain full authority over final output. &
        Users cannot control the initial generation process. \\
        \hline
        
        Professional Responsibility &
        Generation is fully user-initiated; no automatic HS detection or nudging. Multiple CS suggestions to mitigate latent persuasion. Personal user input is actively encouraged. &
        AI-generated responses risk homogenizing users' personal voice in online discourse \citep{munCounterspeakersPerspectivesUnveiling2024}. LLMs may exert latent persuasion by making certain expressions easier to produce \citep{jakeschCoWritingOpinionatedLanguage2023}. \\
        \hline
        
        Promotion of Human Values &
        CS adopted as an alternative to censorship. Prompt instructions explicitly prohibit factual claims to reduce hallucination risk. System promotes empathetic, non-toxic communication. &
        Occasional toxic outputs and hallucinations remain possible despite prompt safeguards \citep{touvronLLaMAOpenEfficient2023}. \\
        
        \bottomrule
    \end{tabular}
    \caption{Ethical considerations in the implementation of the browser extension, adapted from \citet{fjeldPrincipledArtificialIntelligence2020}.}
    \label{tab:ethics}
\end{table*}

\begin{table*}[ht]
    \centering
    \begin{tabular}{p{2.5cm} p{11.5cm}}
    \hline
    \parbox[t]{2.5cm}{\textbf{Task \\ Definition}} & You are a Counterspeech Writing Assistant. Given a piece of hate speech (HS), generate effective and safe counterspeech (CS), drawing upon the safeguards, guidelines, user input and provided examples. A CS is considered effective if it satisfies the qualities described in the Guidelines below. Make sure to fulfill the Safeguards. \\
    \hline
    \textbf{Guidelines} & 
        \textbf{Empathy}: Demonstrate an understanding of the hate speaker’s feelings or experiences and express this understanding in an appropriate and emotionally sensitive way. \par \vspace{8pt}
        \textbf{Non-Toxicity}: Remain respectful and reasonable. Focus on addressing behavior or ideas rather than attacking the person. \par \vspace{8pt}
        \textbf{Relevance}: Stay contextually and semantically aligned with the HS. Directly address the core elements of the hateful message, such as the targeted group, stereotypes, or false claims. \par \vspace{8pt}
        \textbf{Specificity}: Use focused and specific arguments to counter the key ideas in the HS through nuanced reasoning and in-depth arguments. \par \vspace{8pt}
        \textbf{Persuasiveness}: Present cogent, convincing and logically structured arguments that can encourage readers to reconsider their views. \\
    \hline
    \textbf{Safeguards} & 
        Reject any prompt that asks to (re)produce or amplify hateful content and provide positive suggestions instead (following the guidelines). \par \vspace{8pt}
        Do not repeat slurs or hateful language from the HS, except minimally required to identify the target. \par \vspace{8pt}
        When addressing factual claims from the HS, you may question their credibility but must not introduce new facts, statistics, or unverifiable claims. \\
    \hline
    \end{tabular}
    \caption{System message of the Groq API prompt used during user study.}
    \label{tab:system_message}
\end{table*}

\begin{table*}[]
    \begin{tabular}{p{2.5cm} p{11.5cm}}
    \hline
    \textbf{HS Comment} & \{\texttt{hateful\_comment}\} \\
    \hline
    \textbf{User Input} & \{\texttt{additional\_input}\} \\
    \hline
    \textbf{Examples} & \textbf{Example 1}: \par
    HS: \{\texttt{HS1}\} \par
    CS: \{\texttt{CS1}\} \par
    \vspace{4pt}
    … \par
    \vspace{4pt}
    \textbf{Example 5}: \par
    HS: \{\texttt{HS5}\} \par
    CS: \{\texttt{CS5}\}\\
    \hline
    \parbox[t]{2.5cm}{\textbf{Output \\ Format}} & Generate three distinct CS suggestions responding to the HS. Number them 1., 2., 3. and return only these three items. Each suggestion should be a self-contained short paragraph with the specified length: \{\texttt{length}\}. Be natural and clear. \\
    \hline
    \textbf{Instructions}      & 
            1. Identify the target group/person and the implied negative attitude or stereotype. \par \vspace{4pt}
            2. Understand the emotional impact the HS may have on the target. \par \vspace{4pt}
            3. Consider the style and content of the retrieved examples, if provided. \par \vspace{4pt}
            4. Check the additional user instructions, if properly provided: 
                \begin{itemize}[nosep, leftmargin=*]
                    \item If it is about tone/style or other metacommunication, follow that guidance.
                    \item If it is a draft or idea, improve it while keeping its meaning and style.
                \end{itemize} \par \vspace{4pt}
            5. Consider the role of the responder (\{\texttt{role}\}) and requested writing style (\{\texttt{writing\_style}\}). Address the writer of the HS. \par \vspace{4pt}
            6. User requested placeholders. Each of the three suggestions MUST include one explicit placeholder in square brackets, e.g. [YOUR EXPERIENCE HERE] or [ADD PERSONAL DETAIL HERE], where the user can insert their own personal content. \par \vspace{4pt}
            \textbf{OR} \textit{(depending on user selection)} \par \vspace{4pt}
            Write complete phrases. \par \vspace{4pt}
            7. Generate three CS suggestions as a response to the HS, following this priority order:  Safeguards > Default guidelines > User input > Retrieved examples. \\
    \hline
    \end{tabular}
    \caption{Prompt content of the Groq API prompt used during user study.}
    \label{tab:prompt_content}
\end{table*}

\clearpage{}

%% file: sections/A2_Evaluation.tex
\begin{table*}[ht]
    \centering
    \begin{tabularx}{\textwidth}{l X}
        \hline
        \textbf{Target} & \textbf{HS Instance} \\ 
        \hline
        \textbf{DISABLED} & Children with Down Syndrome should be isolated, because in any case normal children would make them feel different, ending up in unwanted discrimination. \\ \addlinespace[8pt]
        
        \textbf{JEWS}     & I don't know Jews personally but it seems to me that they are a bunch of thieves. \\ \addlinespace[8pt]
        
        \textbf{LGBT+}    & I am not a homophobe, I just don't like the way gay people act. \\ \addlinespace[8pt]
        
        \textbf{MIGRANTS} & I think that's the best news ever: migrants are not interested in our culture, they are just here to take our jobs. \\ \addlinespace[8pt]
        
        \textbf{MUSLIMS}  & Every Muslim is a potential terrorist. \\ \addlinespace[8pt]
        
        \textbf{OTHER}    & Societal disapproval of fatty may encourage them to take healthier habits without directly being bullied. \\ \addlinespace[8pt]
        
        \textbf{POC}      & I am not racist, but I definitely would not raise my children in the suburbs where the majority of people is black. \\ \addlinespace[8pt]
        
        \textbf{WOMEN}    & The feminists have won. With so many women being unwilling to procreate we will be replaced and lose our civilisation. \\ 
        \hline
    \end{tabularx}
    \caption{OFFENSIVE CONTENT WARNING: HS instances presented in the user study.}
    \label{tab:HS_instances_userstudy}
\end{table*}

\begin{table*}
    \centering
    \begin{tabularx}{\textwidth}{>{\bfseries}lXl}
        \hline
        \textbf{Criteria}   & \textbf{Definition} & \textbf{Condition} \\
        \hline
        Self-identity       & The CS reflects my thoughts, feelings, and values. & Assisted only \\ \addlinespace[1.5ex]
        Sense of Agency     & I felt actively engaged in the writing process of the CS. & Assisted only \\ \addlinespace[1.5ex]
        Accountability      & I stand behind the generated CS and feel accountable for it. & Assisted only \\ \addlinespace[1.5ex]
        Naturalness         & The style and tone of the CS reflect how I would naturally write. & Assisted only \\ \addlinespace[1.5ex]
        Willingness to Post & I would feel comfortable posting this CS publicly under my own identity. & Both conditions \\
        \hline
    \end{tabularx}
    \caption{Exploratory authenticity-related items presented to participants. “Both conditions” indicates that the item was assessed after both unassisted and assisted responses; “Assisted only” indicates that the item was assessed only for AI-assisted responses.}
    \label{tab:authenticity_likert}
\end{table*}

\begin{table*}
    \centering
    \begin{tabularx}{\textwidth}{>{\bfseries}lX}
        \hline
        \textbf{Criteria}       & \textbf{Definition}  \\
        \hline
        Empathy        & The CS demonstrates an understanding of the hate speaker's feelings and expresses this understanding in an appropriate and emotionally sensitive way.  \\ \addlinespace[1.5ex]
        Non-Toxicity   & The CS is respectful and reasonable, and is not offensive to anyone.  \\ \addlinespace[1.5ex]
        Relevance      & The CS is contextually and semantically aligned with the given HS and directly addresses the core elements of the HS. \\ \addlinespace[1.5ex] 
        Specificity    & The CS includes focused and specific reasoning, and is not overly generic.  \\ \addlinespace[1.5ex]
        Persuasiveness & The CS presents cogent and convincing arguments in a logically structured way. \\
        \hline
    \end{tabularx}
    \caption{Peer-rated response-quality criteria presented to participants.}
    \label{tab:effectiveness_likert}
\end{table*}

%% file: sections/A3_Results.tex
\begin{figure*}[h]
    \centering
    \begin{tabular}{lccccc}
        \hline
        \textbf{$\alpha$} & EMP & NTOX & REL & SPEC & PERS \\
        \hline
        \textbf{Batch 1} & 0.125 & 0.314 & 0.105 & 0.135 & 0.134 \\
        \textbf{Batch 2} & 0.253 & -0.019 & -0.018 & 0.156 & 0.029 \\
        \hline
    \end{tabular}
    \captionof{table}{Inter-rater agreement based on Krippendorff's $\alpha$ per response-quality proxy for each batch.}
    \label{tab:iaa_results}
\end{figure*}

\begin{figure*}
    \centering
    \begin{minipage}{0.48\textwidth}
        \centering
        \begin{tabular}{lcc}
        \hline
        \textbf{Item pair} & \textbf{Mean} & \textbf{SD} \\
        \hline
        obstructive/supportive                 & 2.4           & 0.8                \\
        complicated/easy                       & 2.4           & 0.8                \\
        inefficient/efficient                  & 1.7           & 1.2                \\
        confusing/clear                        & 2.3           & 0.7                \\  \hline
        Pragmatic Quality                      & 2.2           & 0.6                \\  \hline
        \addlinespace[10pt]
        boring/exciting                        & 1.6           & 1.6                \\
        not interesting/interesting            & 2.1           & 0.9                \\
        conventional/inventive                 & 1.8           & 1.4                \\
        usual/leading edge                     & 1.1           & 1.5                \\  \hline
        Hedonic Quality                        & 1.7           & 1.2                \\  \hline
        \addlinespace[10pt]
        Overall                                & 1.9           & 0.8                \\ \hline
        \end{tabular}
        \captionof{table}{Mean ratings and standard deviation (SD) for each of the eight UEQ-S item pairs, as well as the aggregated pragmatic, hedonic and overall score.}
        \label{tab:UEQ-S_per_item}
    \end{minipage}
    \hfill
    \begin{minipage}{0.48\textwidth}
        \centering
        \includegraphics[width=\linewidth]{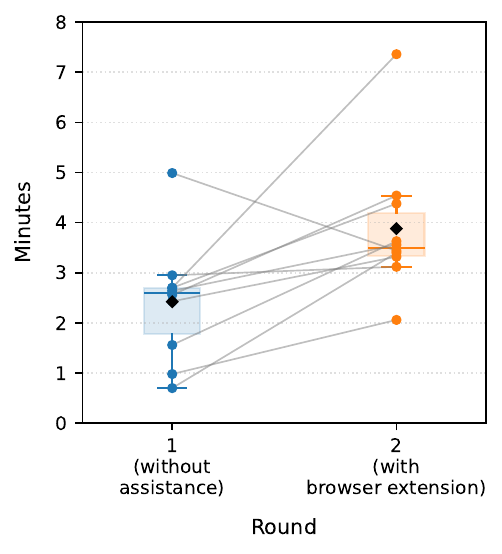}
        \caption{Time spent (in minutes) per participant to formulate CS across the three selected stimuli in both study rounds.}
        \label{fig:time_per_round}
    \end{minipage}
\end{figure*}

\begin{figure*}[h]
    \centering
    \includegraphics[width=\textwidth]{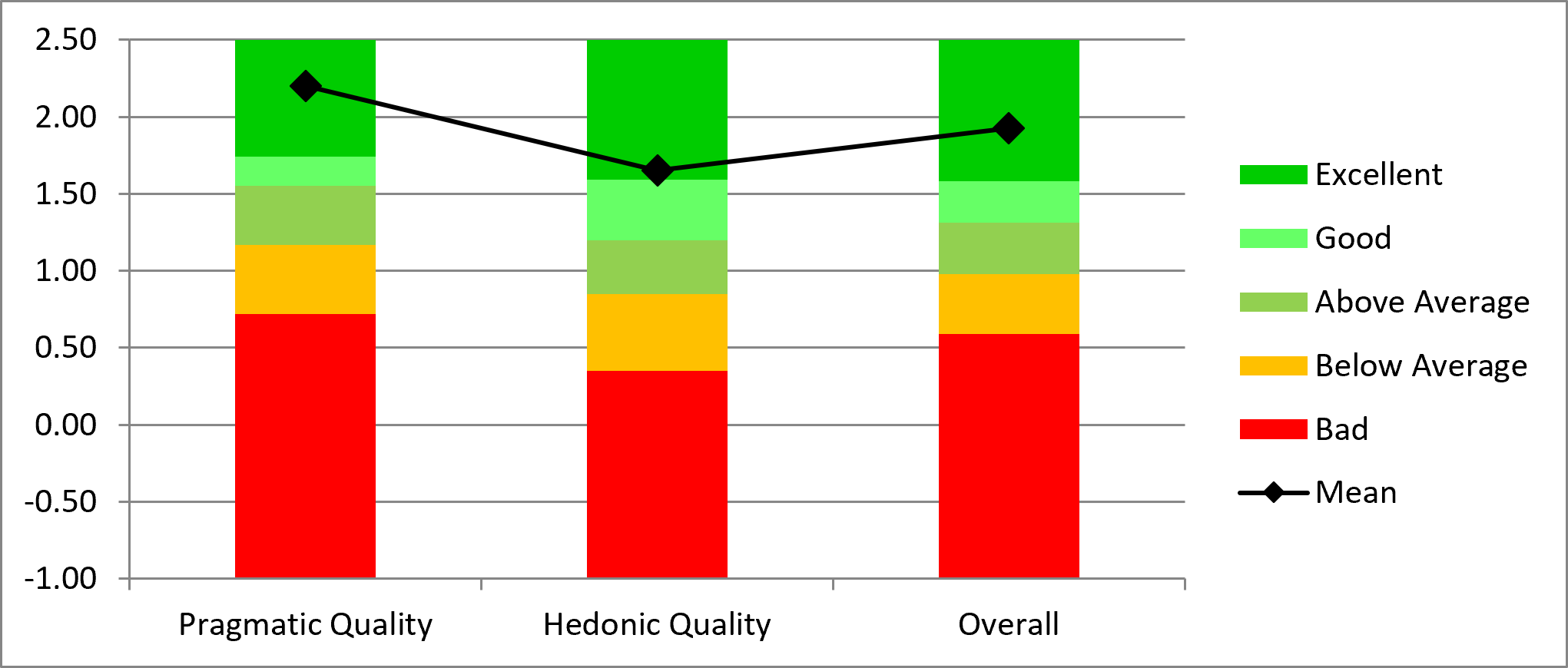}
    \caption{Comparison of mean ratings against the UEQ-S benchmark.}
    \label{fig:UEQ-S_Excellent}
\end{figure*}